\newif\ifPDF
\ifx\pdfoutput\undifine\PDFfalse
\else\ifnum\pdfoutput >0\PDFtrue
\else\PDFfalse
\fi
\fi

\ifPDF
   \documentclass[10pt,twocolumn,twoside,a4paper,pdftex]{article}
\else
   \documentclass[10pt,twocolumn,twoside,a4paper,dvips]{article}
\fi

\usepackage[small]{caption}
\def\abstract{\begin{center}
{\it \large Abstract\vspace{-.5em}\vspace{0pt}}
\end{center}}

\usepackage{amsfonts}
\usepackage{amssymb}
\usepackage{verbatim}
\usepackage{graphicx}
\usepackage{hyperref}

\ifPDF
   \hypersetup{%
     pdftoolbar=treu,%
     pdfmenubar=true}
\else
   \usepackage{epsfig}
   \usepackage{pslatex}
\fi

\newcommand{\ssection}[1]{%
     \section[#1]{\centering \sc #1}}

\def\abstract{\begin{center}
{\it \large Abstract\vspace{-.5em}\vspace{0pt}}
\end{center}}

\title
{ \vspace{-1cm} 
\Large The IceCube Project}
\author{\large C. Spiering for the IceCube Collaboration
\footnote{a full author list is given at the end of this paper}\\
\\ \normalsize DESY Zeuthen, Platanenallee 6, D15739 Zeuthen, Germany\\
\normalsize christian.spiering@desy.de}

\begin{document}
\date{}
\maketitle

\thispagestyle{empty}
\pagestyle{empty}

\begin{abstract}
This talk gives a brief description of goals, design,
expected performance and status of the IceCube project.
\end{abstract}

\ssection{Physics Goals}

The main goal of the IceCube project \cite{IceCube} is to
extend the region of the Universe explored by neutrinos and
thereby to test fundamental laws of physics, to obtain a
different view of astronomical objects, and to learn about the
origin of the highest-energy cosmic rays. 

Science topics
include the search for steady and variable sources of high
energy neutrinos like Active Galactic Nuclei (AGN), Supernova
Remnants (SNR) or microquasars, as well as the search
for neutrinos from burst-like sources like 
Gamma Ray Bursts (GRB). The sensitivity of
IceCube to astrophysical sources of high energy muon neutrinos
is described in \cite{IceCubeapp} and in Section V.

Similar to the Mediterranean projects discussed at this workshop,
IceCube can also tackle a series of questions 
beside high energy neutrino astronomy. They include
the search for neutrinos from
the decay of dark matter particles (WIMPs) and the search
for magnetic monopoles or other exotic particles like strange 
quark matter or the Q-balls predicted by SUSY models
(see for reviews \cite{CSICRC, Halzen}).

There are, however, two modes of operation which are 
not -- or nearly not  --
possible for detectors in natural water.
Firstly, due to the low light activity of the
surrounding medium, the PMT counting rate is below
1 kHz. This enables the detection of feeble rate increases
as caused, for instance, by interactions of Supernova
burst neutrinos. IceCube can monitor the full Galaxy for MeV neutrinos
from Supernova explosions. Secondly, IceCube can be
operated in coincidence with a surface air shower array, IceTop.
This allows to study questions like the chemical composition of 
cosmic rays up to $10^{18}$ eV, to calibrate IceCube, and
to use IceTop as a veto for background rejection.

\ssection{Detector Configuration}

The configuration of IceCube is shown in Fig.~1. IceCube is an
array of 4800 optical modules (OMs) on 80 strings, regularly
spaced by 125 m. It covers an area of approximately 1 km$^2$,
with the OMs at depths of 1.4 to 2.4 km below surface. Each string
carries 60 OMs, vertically spaced by 17 m. The strings are
arranged in a triangular pattern.  At each hole, one
station of the IceTop air shower array will be positioned.
An IceTop station consists of two ice tanks of total area
7 m$^2$. Two of these tanks are being installed during the
current (2003/04) season.

\vspace{-3mm}
\begin{figure}[h,t]      
\begin{center}
\includegraphics[width=.39\textwidth,angle=-90]{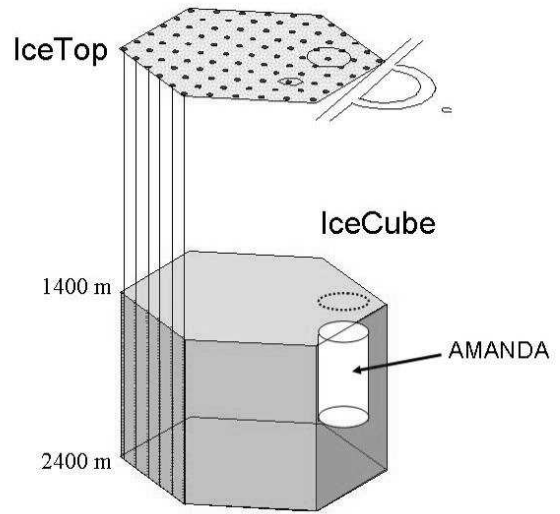}
\caption{
IceCube, IceTop and Amanda}
\label{fig:icecube}
\end{center}
\end{figure}

\vspace{-4mm}

The present Amanda-II detector 
\cite{Bernardini} will be integrated into IceCube. 
The present runway as well as 
the old  South Pole station, buried not far from
Amanda under the snow, prevent an extension of IceCube into
the corresponding  area and therefore a central location of Amanda
within IceCube. Still, IceCube will deliver efficient
veto information for low energy cascade-like events 
or short horizontal tracks recorded in Amanda. Horizontal
tracks could  be related to neutrinos steming from WIMP 
annihilations in the Sun.

Various configurations with the number of OMs ranging from
2400 (half the design number) to 9600 (twice the design number),
with equally spaced strings and with nested subarrays of larger
density, and also a variety of ``exotic'' configurations have been
studied in detail \cite{Leuthold}. 
The present configuration is tailored for best
sensitivity to muon neutrinos in the energy range of TeV-100 TeV.
Better sensitivity at low energy may
be obtained by improved sensitivity of individual OMs 
(i.e. by application of wave length shifters \cite{resconi}, see below). 
Better sensitivity at higher energies as obtained by 
{\it additional} strings along sparsely equipped circle(s)
around IceCube is discussed in \cite{HalzenHooper}.

\ssection{The Digital Optical Module}

The IceCube optical module is sketched in Figure 2. It contains 
a 10-inch diameter PMT HAMAMATSU R-7081. The main argument
for this choice was, apart from excellent charge and time
resolution, the low noise of a only few hundred Hz (at temperatures
of -20 to -40 $^o$C). Low dark noise of the PMTs
is not a strong criterion
for water detectors since in natural water 
the ambient noise dominates. In
ice, PMTs can be operated without tight local
coincidences if only their noise is smaller than a few kHz.
A compromise with respect to noise would
significantly deteriorate the
performance, in particular for detection of  low-energy
Supernova neutrinos. The PMT is embedded in a transparent gel
and shielded against the Earth's magnetic
field by a mu-metal grid.

\medskip

\begin{figure}[h,t]      
\begin{center}
\includegraphics[width=.45\textwidth]{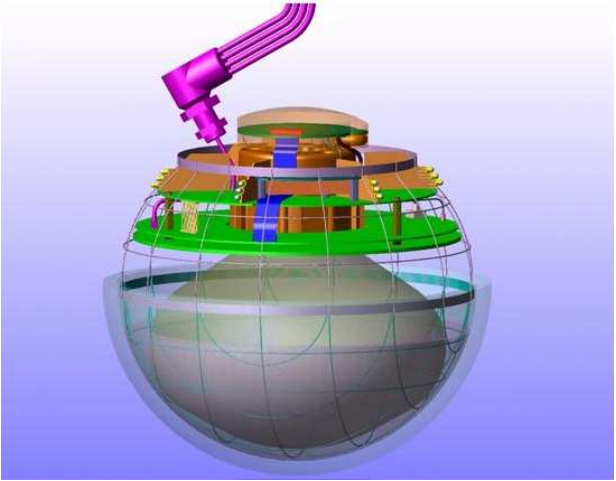}
\caption{
Schematic view of the IceCube DOM}
\label{fig:OM}
\end{center}
\end{figure}

The PMT anode signal (gain $\sim 5 \cdot 10^7$) is digitized
within the OM (or Digital Optical Module, DOM)  
and sent to surface via electrical 
twisted-pair cables, one twisted pair for two DOMs. The DOM
contains several electronic components: the signal processing board,
a LED flasher board for calibration purposes, and the
PMT base with the high voltage power supply.

The requirements for time resolution and dynamic range are:

\begin{itemize}
\item waveform recorded with 250 MHz over the first 0.5 $\mu$s
and 40 MHz over 5 $\mu$s.
\item each pulse time stamped with 7 ns r.m.s.
\item dynamic range 200 photoelectrons over first 15 ns and
2000 photoelectrons when integrated over 5 $\mu$s.
\item dead time $< 1 \%$
\item noise rate $<$ 500 Hz.
\end{itemize}

The fine sampling is done with 
the Analog Transient Waveform Recorder (ATWR), an ASIC with four
channels, each capable to capture 128 samples with 200-800 Hz.
The 40 MHz sampling is performed by a commercial FADC.

\begin{figure}[h,t]      
\begin{center}
\includegraphics[width=.42\textwidth]{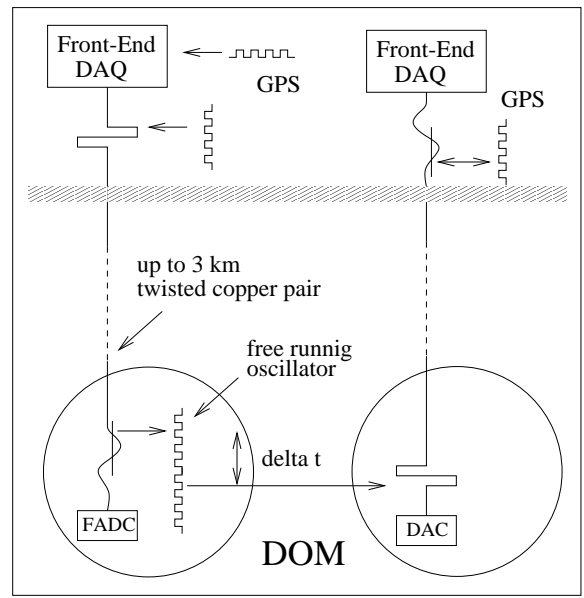}
\caption{
Principle of time calibration and offset determination}
\label{fig:timing}
\end{center}
\end{figure}

The effects of light scattering on the photon
arrival time are dominant compared to effects of PMT jitter
and time calibration. It has been shown that
reconstruction quality only worsens if
the jitter increases beyond
10-15 ns, resulting in a design value of 7 ns
for overall timing accuracy.
Time calibration over 2-3 km electrical cable 
with a few-nanosecond accuracy is a challenge. In IceCube it is 
solved by sending a bipolar signal 
%at known intervals
from the DAQ
to the DOM (see Fig.\ref{fig:timing}, left). 
The leading edge of this signal is synchronized
to the surface clock common to all DOMs. The signal is
considerably dispersed over 2 km cable and is
subject to baseline variations and noise. Therefore
the calibration signal arriving at the DOM is 
digitized by a FADC. 
The full waveform information allows for
a correction with respect to the mentioned effects.
%and indeed results in
%a synchronization of the local clock in each DOM with
%an accuracy of $< 5$ nsec. 
Since the stability
of the local oscillator in the DOM
is better than $10^{-10}$, the 
calibration process has to be repeated only every
10 seconds.

The time offset is determined by a method
known as Reciprocal Active Pulsing (RAP) \cite{RAP}.
%The principle is sketched in Fig.~\ref{fig:timing}.
In response to the calibration signal, 
after a well-defined delay $\delta t$,  the same bipolar
signal is sent back to surface and treated there in exactly
the same way as the downward calibration signal had been
treated in the DOM. This procedure yields 
an overall delay $T_{roundtrip}$.
With the same signal treatment at surface and in the DOM,
one gets $T_{up} = T_{down} = 0.5 (T_{roundtrip} - \delta t$).
This information defines the time offset for each DOM and can
be determined by multiple round trip measurements 
with better than 1 ns accuracy. Over all, the
instantaneous accuracy of the time calibration is
expected to be 5 ns r.m.s.

\ssection{Drilling and deployment}

The drill technology for IceCube is well understood from experience
with several years of Amanda deployments. Holes of 60 cm
diameter are drilled with 80 $^o$C hot water. For IceCube,
a new drilling system (Enhanced Hot Water Drill, EHWD)
has been constructed. The power for heaters and pumps
of the EHWD will be $\sim$ 5 MW, compared to 2 MW for Amanda.
This, and larger diameter and length of the water transporting
hoses, will result in only 40 hours needed to drill a
2400 m deep hole (three times faster than with the old
Amanda drill). The fuel consumption is reduced by one third.

Mounting, testing and drop of a string with 60 DOMs 
is expected to take about 20 hours. Since the set-up of the
drilling and deployment system at the beginning of the
Antarctic summer season will be reduced from 5 
to 3 weeks, deployment of 16 strings per season
is feasible.
 
\ssection{Physics Performance}

In this section, we  summarize
results on {\it track} detection at TeV to PeV energies published in 
\cite{IceCubeapp}. IceCube performance at higher
energies is considered in \cite{Yoshida}.

\begin{figure}[h,t]
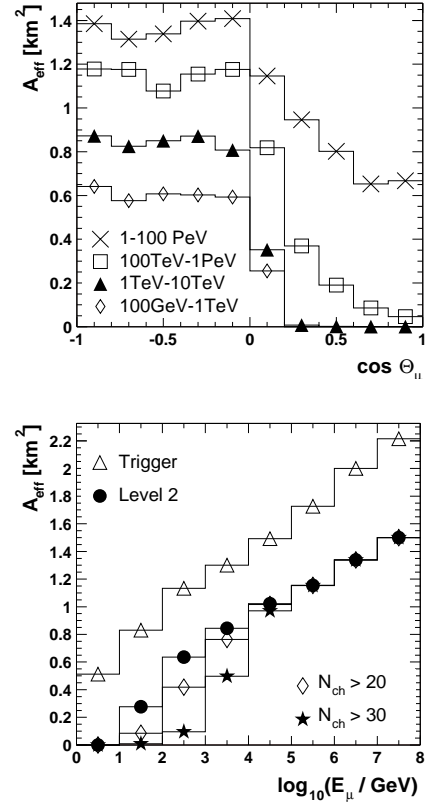
      
\begin{center}
\includegraphics[width=.30\textwidth]{aeff.epsi} \\
\vspace{5mm}
\includegraphics[width=.30\textwidth]{aeff_endep.epsi}
\caption{
Effective area for muon as a function of zenith angle (top)
and as a function of muon energy (bottom).
The lower figures refers to muons from the
lower hemisphere ($\cos \theta < 0$). See text
for further explanations.
}
\label{fig:area}
\end{center}
\end{figure}

Figure \ref{fig:area} (top) shows the effective 
area after $\sim 10^{-6}$ reduction 
of events from downward muons
as a function of the muon zenith angle. Whereas at 
TeV energies IceCube is blind towards the upper hemisphere, 
at PeV and beyond the aperture extends above the horizon and 
allows observation of the Southern sky. Figure 
\ref{fig:area} (bottom) shows the effective area as function
of the muon energy at the detector, averaged
over the lower hemisphere. The upper curves refer to events passing 
trigger criteria (triangles) and fake event reduction (full circles). 
The other two curves show the effective area after tuning the cuts for 
best sensitivity to  hypothetical steady point source fluxes which 
follow $E^{-2}$ (stars) and  $E^{-2.5}$ (diamonds)
power laws, respectively.  The
threshold is around 1 TeV for  $E^{-2}$ spectra. The
energy-dependent optimum cut is applied to $N_{ch}$, 
the number of fired PMTs.

\begin{figure}[h,t]      
\begin{center}
\includegraphics[width=.30\textwidth]{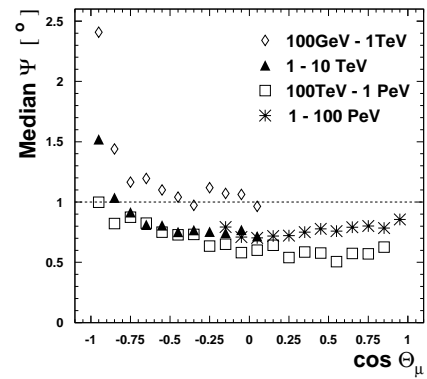}
\caption{
Pointing resolution (median space angle)
for neutrino-induced muon events
as a function of zenith angle}
\label{fig:pointing}
\end{center}
\end{figure}

Fig.~\ref{fig:pointing} 
shows the pointing resolution for neutrino-induced muon 
events after background rejection and assuming an $E^{-2}$ spectrum. 
For not too steep angles 
the resolution is 0.6 to 0.8 degrees, improving with energy. We expect that
evaluation of waveform information will improve these numbers significantly,
at least at high energies. Paradoxically, the reason 
is the strong light scattering which is known to be a clear
draw-back with respect to the accuracy of first-photon
arrival times. Part of this drawback can likely be turned into an
advantage since scattering 
modifies the arrival time distribution as a function of 
distance between source and receiver. A wider distribution 
of a many-photoelectron signal recorded 
by a single PMT indicates a more distant source \cite{wiebusch}.
This additional information can be used to improve reconstruction, 
provided the
amount of light is high enough to generate multi-photoelectron waveforms in
many PMTs. However, a resolution as good as projected for water detectors 
seems hardly achievable.

Fig.~6 shows the energy spectra of selected neutrinos for $E^{-2}$ sources.
(Note that Fig.~4 refers to muon energy, but Fig.~6 to neutrino energy).
The top part shows the spectrum after fake event rejection and after
cuts tailored to get best sensitivity to {\it point sources}. 
It confirms the TeV threshold demonstrated in
Fig.~4. The bottom part shows the spectrum if the cuts are optimized for best
sensitivity to a {\it diffuse} $E^{-2}$  flux. Clearly the large 
background from
atmospheric neutrinos over $2 \pi$ or more requires harder energy cuts than
in the case of point sources. The threshold is now about 100 
times higher than for the point source analysis.

\begin{figure}[h,t]
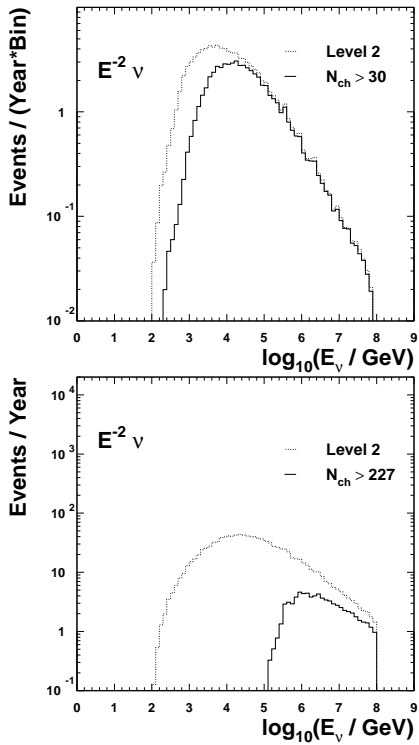
      
\begin{center}
\includegraphics[width=.30\textwidth]{passpoint.epsi}
\includegraphics[width=.30\textwidth]{passdiff.epsi}
\caption{
Energy spectra of selected neutrinos for a $E^{-2}$ source.
Cuts have been optimized to get the best sensitivity to
point sources (top) and to a diffuse flux (bottom).
The cutoff at $10^8$ GeV is due to the limited 
energy range of the simulation.}
\label{fig:spectra}
\end{center}
\end{figure}

\vspace{-2mm}
Fig.~7 shows the expected sensitivity to diffuse fluxes as 
function of neutrino energy.
Solid lines indicate the expected 90\% c.l. limits for  
$E^{-2}$ and $E^{-1}$ spectra, respectively, calculated for a data 
taking period of three years. The lines extend over the energy range
containing 90\% of the expected signal. The dashed lines indicate the 
Stecker and Salamon model for photo-hadronic interactions in AGN 
cores \cite{SS}. The dotted
line corresponds to the Mannheim, Protheroe and Rachen model
on neutrino emission from photo-hadronic interactions in AGN 
jets \cite{MPR}. In case of
no signal, these models could be rejected with 
model rejection factor (mrf) \cite{Hill} of
$10^{-3}$ and $10^{-2}$ respectively. Also shown is the GRB
estimate by Waxman and Bahcall \cite{WB}
%which could be tested using about hundred bursts.
which would yield of the order of ten events coinciding with
a GRB, for 1000 GRBs monitored.

\begin{figure}[h,t]      
\begin{center}
\includegraphics[width=.45\textwidth]{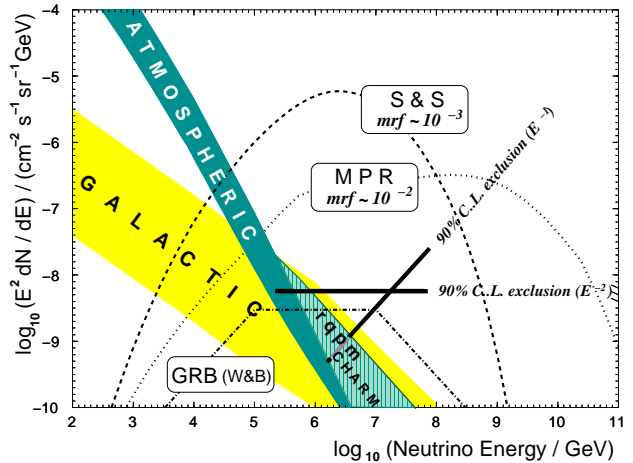}
\caption{
Expected sensitivities of the IceCube detector. See text for
explanations.}
\label{fig:sensitivity}
\end{center}
\end{figure}

%\vspace{-2mm}
Apart from tracks, IceCube can map {\it cascades}, with an energy
resolution of about 30\% at high energies.
Compared to Amanda,  IceCube  has
a much larger central volume shielded by outer "veto layers". 
This allows
significantly better recognition of isolated cascades -- 
an important issue
for the detection of  $\nu_e$ or $\nu_{\tau}$ interactions.

\ssection{Possible extensions}

Parallel to the implementation of the IceCube baseline design,
methods are studied to extend the capabilities of the
South Pole detector towards higher as well as towards 
lower energies. With a volume
of a cubic kilometer, IceCube will surpass Amanda by a factor of
40 (and Super-K by a factor of 1000). These numbers refer to the
sensitivity in the energy range of TeV-PeV. 

For energies below
a few hundred GeV the light emission is too low to fire
many PMTs, and the effective volume decreases
drastically. 
In order
to increase the sensitivity at lower energies, more light
has to be collected. First encouraging result have been
obtained in production of transparent hats for the
glass spheres housing the PMTs. These hats are doped with
wavelength shifter which moves the light from wavelengths
below 300 nm to wavelengths above 320 nm where the spheres 
are becoming transparent \cite{resconi}.
The resulting increase in light collection may be as high
as 40\%, following first measurements and estimates. 
This would result in
a lower energy threshold and better reconstruction.

For energies above 100 PeV, on the other
hand, expected neutrino fluxes are so small that even a cubic
kilometer is not sufficient to catch a few events.
Instead, volumes of ten or hundred cubic kilometers are required.
With the given spacing of PMTs (17 m
vertically, 125 m horizontally) this is not affordable, neither
logistically nor financially. A larger volume could be
reached by larger PMT spacing
(see e.g. \cite{HalzenHooper}. The increase of distance, however,
is limited by the absorption length of light ($\sim$100 m)
and the smaller amount of light reaching the PMTs. 
In order to allow a much larger spacing, an information carrier
with smaller absorption in ice than light has to be used. Radio
wave detection has been successfully applied to derive
ultra-high energy flux limits with the RICE experiment,
located above the Amanda experiment, at a depth of a few hundred 
meters \cite{Rice}. Another  method is the
detection of acoustic waves generated in high-energy
particle cascades. The attenuation length of this signal
in pure ice at -40$^o$C is about 2 km \cite{Price}.
R\&D work is underway to develop sensitive
acoustic sensors with good signal-to-noise behaviour, to
test them at accelerators, and to understand noise
behaviour in open reservoirs of water and ice
\cite{acoustic}. In-ice tests
will show how well the method will work in natural ice.

\vspace{-1mm}
\ssection{Status and outlook}

The IceCube collaboration includes about 150 scientists from
institutions in Belgium, Germany, Japan, Netherlands, New Zealand,
Sweden, UK, USA and Venezuela. In the USA IceCube is handled
as MRE (Major Research Equipment) project of the National Science
Foundation, NSF. Funding of the start-up phase started in
FY 2002. In the mean time, the project moved from the start-up
phase to the implementation phase and is now fully installed
as a MRE project.
Apart from the US budget, significant funding 
has been approved in Belgium, Germany and Sweden.

The present plan foresees transportation of the EHWD to the Pole
at the end of 2004, and drilling of a few first holes in January
2005. Amanda will be integrated into the year-by-year
increasing IceCube. The growing detector will take data
during construction, with each string coming online within days 
after deployment. Construction of IceCube is to be completed in 2010,
followed by $\sim$ 10 years of data taking.

\section*{IceCube Author List}

\vspace{-1mm}
\begin{sloppypar}
\noindent
{\scriptsize
A.~Achterberg$^{25}$, 
M.~Ackermann$^{6}$
J.~Ahrens$^{15}$, 
J.N.~Bahcall$^{8}$, 
X.~Bai$^{1}$, 
R.C.~Bay$^{14}$, 
T.~Becka$^{15}$, 
K.-H.~Becker$^{2}$, 
J.~Bergmans$^{25}$, 
D.~Berley$^{16}$
% ...
}

\end{sloppypar}

\vspace*{0.2cm} 

{\footnotesize
\noindent
   (1) Bartol Research Institute, University of Delaware, Newark, DE 19716, USA
   \newline
   (2) Fachbereich 8 Physik, BUGH Wuppertal, D-42097 Wuppertal, Germany
   \newline
   (3) Universit\'e Libre de Bruxelles, Science Faculty CP230, Boulevard du Triomphe, B-1050 Brussels, Belgium
   \newline
   (4) CTSPS, Clark-Atlanta University, Atlanta, GA 30314, USA
   \newline
   (5) Dept. of Physics, Chiba University, Chiba 263-8522 Japan
   \newline
   (6) DESY-Zeuthen, D-15738 Zeuthen, Germany
   \newline
   (7) Blackett Laboratory, Imperial College, London SW7 2BW, UK
   \newline
   (8) Institute for Advanced Study, Princeton, NJ 08540, USA
   \newline
   (9) Dept. of Physics and Astronomy, University of Kansas, Lawrence, KS 66045, USA
   \newline
   (10) Lawrence Berkeley National Laboratory, Berkeley, CA 94720, USA
   \newline
   (11) Dept. of Astronomy and Astrophysics, Pennsylvania State University, University Park, PA 16802, USA
   \newline
   (12) Dept. of Physics, Pennsylvania State University, University Park, PA 16802, USA
   \newline
   (13) Dept. of Physics, Southern University, Baton Rouge, LA 70813, USA
   \newline
   (14) Dept. of Physics, University of California, Berkeley, CA 94720, USA
   \newline
   (15) Institute of Physics, University of Mainz, Staudinger Weg 7, D-55099 Mainz, Germany
   \newline
   (16) Dept. of Physics, University of Maryland, College Park, MD 20742, USA
   \newline
   (17) Dept. of Physics, George Mason University, Fairfax, VA 22030, USA
   \newline
   (18) University of Mons-Hainaut, 7000 Mons, Belgium
   \newline
   (19) Departamento de F\'{\i}sica, Universidad Sim\'on Bol\'{\i}var, Caracas, 1080, Venezuela
   \newline
   (20) Dept. of Astronomy, University of Wisconsin, Madison, WI 53706, USA
   \newline
   (21) Dept. of Physics, University of Wisconsin, Madison, WI 53706, USA
   \newline
   (22) SSEC, University of Wisconsin, Madison, WI 53706, USA
   \newline
   (23) Physics Dept., University of Wisconsin, River Falls, WI 54022, USA
   \newline
   (24) Division of High Energy Physics, Uppsala University, S-75121 Uppsala, Sweden
   \newline
   (25) Faculty of Physics and Astronomy, Utrecht University, NL-3584 CC Utrecht, The Netherlands
   \newline
   (26) Dept. of Physics, Stockholm University, SE-10691 Stockholm, Sweden
   \newline
   (27) Vrije Universiteit Brussel, Dienst ELEM, B-1050 Brussels, Belgium
   \newline

}

\end{document}